\documentclass[sigconf]{acmart}

\usepackage{adjustbox}
\usepackage{multirow}
\usepackage{pifont}
\usepackage{colortbl}
\usepackage{tabularx}
\usepackage{enumitem}
\usepackage{tcolorbox}
\usepackage[absolute,overlay]{textpos}

\AtBeginDocument{%
  }

\setcopyright{acmlicensed}
\copyrightyear{2018}
\acmYear{2018}
\acmDOI{XXXXXXX.XXXXXXX}
\acmConference[Conference acronym 'XX]{Make sure to enter the correct
  conference title from your rights confirmation email}{June 03--05,
  2018}{Woodstock, NY}
\acmISBN{978-1-4503-XXXX-X/2018/06}

\begin{document}


\begin{textblock*}{\paperwidth}(1.5cm, 1.2cm) 
    \includegraphics[height=1.5cm]{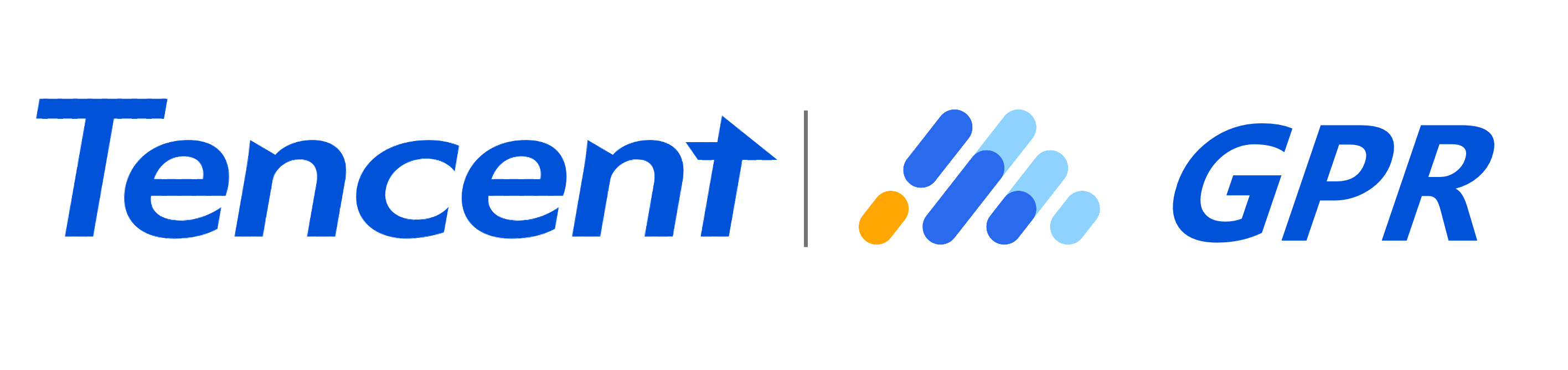} 
\end{textblock*}


\settopmatter{printacmref=false}

\newcommand{\stitle}[1]{\vspace*{0.5em}\noindent{\bf #1\/}}
\newcommand{\squishlist}{
	\begin{list}{$\bullet$}
		{ \setlength{\itemsep}{1pt}
			\setlength{\parsep}{1pt}
			\setlength{\topsep}{2.5pt}
			\setlength{\partopsep}{0.5pt}
			\setlength{\leftmargin}{1em}
			\setlength{\labelwidth}{1em}
			\setlength{\labelsep}{0.6em}
		}
	}
	\newcommand{\squishend}{
	\end{list}
}

\title{End-to-End Semantic ID Generation for Generative Advertisement Recommendation}

\author{{\Large Jie Jiang $^{1*}$, Xinxun Zhang $^{2*}$, Enming Zhang $^{1*}$, Yuling Xiong $^{1*}$, Jun Zhang $^1$, Jingwen Wang $^1$, Huan Yu $^1$, Yuxiang Wang $^2$, Hao Wang $^2$, Xiao Yan $^2$, Jiawei Jiang $^{2\dagger}$}}

\affiliation{
  \institution{$^1$ Tencent Inc., China; $^2$ Wuhan University, China}
  \country{}
}
\email{{zeus, buzzzhang, whitnyxiong, neoxzhang, joywinwang}@tencent.com}
\email{{xxzhangstu, nai.yxwang, wanghao.cs, yanxiaosunny, jiawei.jiang}@whu.edu.cn}



\renewcommand{\shortauthors}{Jiang et al.}

\begin{abstract}
  Generative Recommendation (GR) has excelled by framing recommendation as next-token prediction. This paradigm relies on Semantic IDs (SIDs) to tokenize large-scale items into discrete sequences. Existing GR approaches predominantly generate SIDs via Residual Quantization (RQ), where items are encoded into embeddings and then quantized to discrete SIDs. However, this paradigm suffers from inherent limitations: 1) Objective misalignment and semantic degradation stemming from the two-stage compression; 2) Error accumulation inherent in the structure of RQ. To address these limitations, we propose UniSID, a \textbf{Uni}fied \textbf{SID} generation framework for generative advertisement recommendation. Specifically, we jointly optimize embeddings and SIDs in an end-to-end manner from raw advertising data, enabling semantic information to flow directly into the SID space and thus addressing the inherent limitations of the two-stage cascading compression paradigm. To capture fine-grained semantics, a multi-granularity contrastive learning strategy is introduced to align distinct items across SID levels. Finally, a summary-based ad reconstruction mechanism is proposed to encourage SIDs to capture high-level semantic information that is not explicitly present in advertising contexts. Experiments demonstrate that UniSID consistently outperforms state-of-the-art SID generation methods, yielding up to a 4.62\% improvement in Hit Rate metrics across downstream advertising scenarios compared to the strongest baseline.

\end{abstract}



\maketitle

\def\thefootnote{*}\footnotetext{These authors contributed equally to this work}
\def\thefootnote{$\dagger$}\footnotetext{Corresponding author}
\section{Introduction}
Driven by the huge success of large language models (LLMs) across diverse domains \cite{achiam2023gpt,touvron2023llama,zhou2024large}, recommender systems have increasingly shifted toward generative modeling \cite{li2024survey,zhai2024actions}. In contrast to conventional deep learning-based recommenders that rely on multi-stage cascades or funnel-style pipelines, generative recommendation (GR) casts recommendation as next-token prediction and directly generates the next item a user is likely to interact with \cite{zhang2025gpr,zhou2025onerec,han2025mtgr}. This formulation has demonstrated strong empirical performance in real-world applications, including e-commerce recommendation \cite{yi2025recgpt}, search recommendation \cite{wang2025nezha}, advertising \cite{zhang2025gpr}, and video recommendation \cite{zhou2025onerec2}, thereby providing a unified and scalable approach to sequential user modeling.

\begin{figure}[t]
\centering
\includegraphics[width=0.95\linewidth]{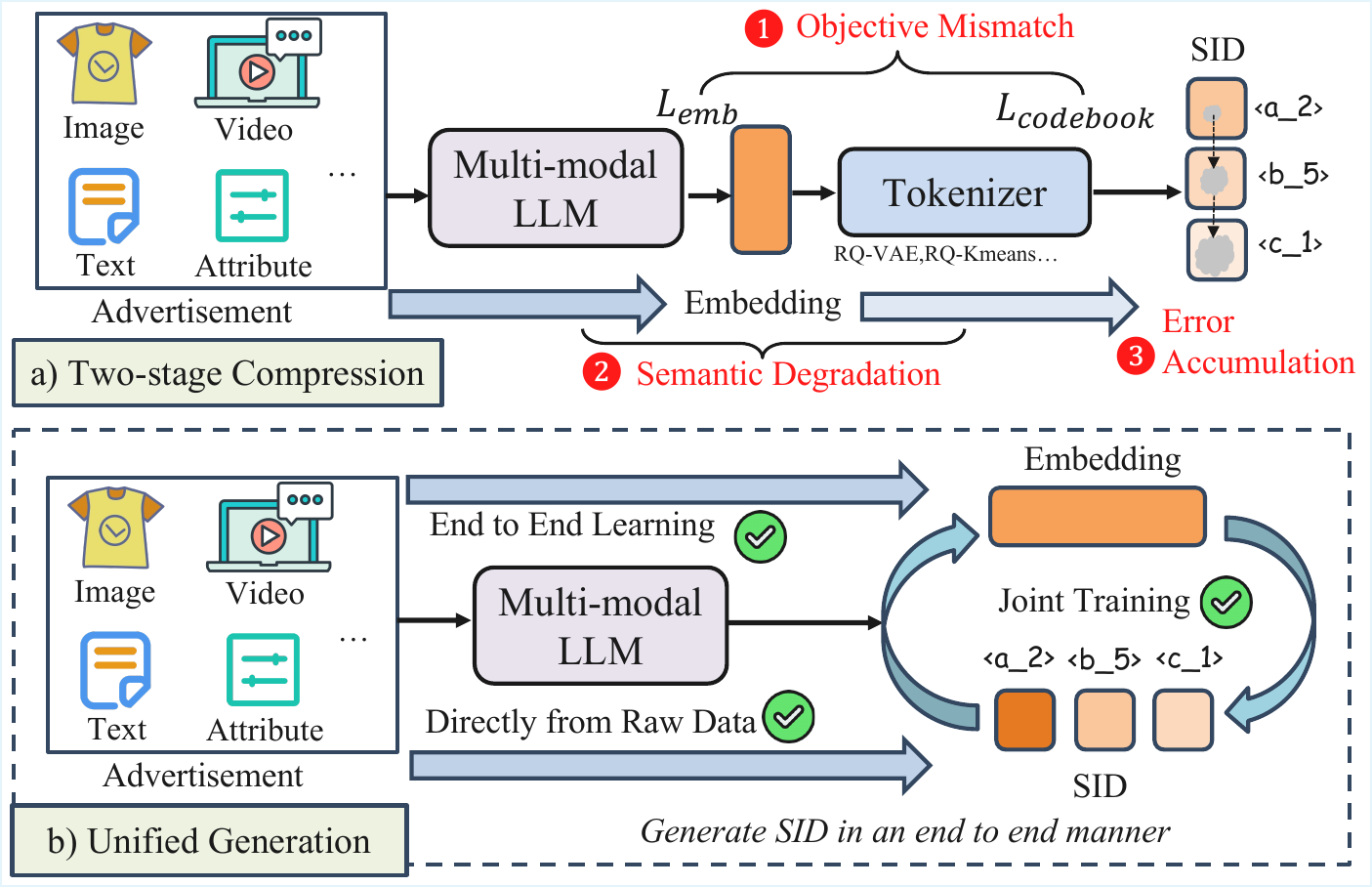} 
\caption{Two-stage cascaded compression of current methods and unified generation of SID of our method.}
\label{fig_motivation}    
\end{figure}

Semantic IDs (SIDs) are a key enabler of GR, mapping billions of items into compact sequences of discrete tokens \cite{hou2023learningsid}. By compressing the item space while preserving compatibility with next-token prediction, SIDs substantially improve the efficiency and scalability of GR \cite{rajput2023recommender,li2025survey}.
As shown in Figure~\ref{fig_motivation}(a), most existing SID generation methods adopt a two-stage paradigm built upon Residual Quantization (RQ): items are first encoded into dense embeddings and subsequently discretized into token sequences \cite{rajput2023recommender,zhou2025onerec,zhang2025gpr,ye2025align,hou2025generating}. 
For instance, OneRec \cite{zhou2025onerec} leverages RQ-Kmeans \cite{luo2025qarm} to unify video GR modeling, whereas GPR \cite{zhang2025gpr} utilizes RQ-Kmeans+ for end-to-end optimization in advertising GR. 
Nevertheless, these approaches still hinge on the two-stage RQ pipeline, in which SID construction is conditioned on pre-trained embeddings rather than being learned end-to-end from raw item features.
Consequently, this two-stage cascading compression design ultimately constrains the modeling capacity of GR for three reasons.

We highlight three limitations of the prevailing paradigm:
\noindent\begin{itemize}[leftmargin=*]
\item[\ding{182}] \textit{Objective misalignment.} Decoupled training objectives across stages induce an inherent optimization mismatch: the embedding-learning stage is trained to produce semantically rich item embeddings, whereas the subsequent SID generation stage is optimized to emit discrete tokens amenable to next-token prediction. This inconsistency precludes end-to-end co-optimization toward a unified objective, yielding suboptimal SID representations.
\item[\ding{183}] \textit{Semantic degradation.} The cascaded pipeline generates SIDs solely from pre-trained embeddings, thus the SID stage cannot directly utilize raw item features (e.g., multimodal/attribute signals) or adapt representations to the discrete code space. This bottleneck can discard critical semantics and degrade SID fidelity.
\item[\ding{184}] \textit{Error accumulation.} RQ-based SID generation hierarchically quantizes item embeddings to approximate fine-grained semantics. However, this hierarchy introduces compounding errors. Quantization noise accumulates across levels, and each level observes only the residual from the previous stage, thereby rendering the available information progressively sparser toward deeper layers. Therefore, SID at later levels tends to be noisier and less reliable.
\end{itemize}
The cumulative effect of these limitations hinders the effectiveness of existing two-stage compression
approaches for high-quality SID generation in GR. This naturally raises the following question:

\begin{quote}
\textit{Can we design a unified end-to-end SID generation framework that breaks away from the two-stage cascading compression paradigm?}
\end{quote}
\stitle{Our Solution.}
Motivated by these observations, we propose UniSID, a \textbf{Uni}fied \textbf{SID} generation framework for generative advertisement recommendation.
To address objective misalignment, UniSID replaces the decoupled two-stage pipeline with a single end-to-end training objective that jointly learns SIDs and embeddings directly from raw advertising data.
To address semantic degradation, we introduce an advertisement-enhanced input schema that linearizes heterogeneous advertisement signals (e.g., task instructions, images, text, and structured attributes) into a unified token sequence, and then appends learnable SID tokens together with an embedding token.
By directly injecting raw multimodal and attribute semantics into the SID space, this design bypasses the pre-trained embedding bottleneck and mitigates semantic loss induced by cascaded compression.
To address Error accumulation in hierarchical RQ, UniSID further avoids layer-wise residual compression.
Instead, each SID layer is predicted from the same full contextual advertisement, ensuring that all layers access complete advertising information and alleviating progressive information sparsification.

Building on this unified pipeline, we introduce two semantics-preserving objectives that further curb semantic degradation and error accumulation while remaining compatible with end-to-end optimization.
First, a multi-granularity contrastive learning strategy enforces granularity-specific semantic consistency by constructing SID-level positive pairs, explicitly regularizing each SID layer to be semantically faithful.
Second, a summary-based advertisement reconstruction mechanism distills advertisement attributes into high-level semantics and reconstructs them from SIDs, encouraging SIDs to preserve key information that may be implicit in raw advertising contexts and providing an auxiliary supervision signal complementary to the unified objective.

We conduct a comprehensive evaluation of UniSID across diverse tasks, including SID quality, next-advertisement prediction, and advertisement retrieval in industrial advertising scenarios, alongside next-item prediction on public benchmarks. Experimental results demonstrate that UniSID consistently outperforms SOTA baselines, achieving maximum improvements of 2.14\% on SID quality, 4.01\% on next-advertisement prediction, 45.46\% on advertisement retrieval, and 11.83\% on next-item prediction. Furthermore, ablation studies validate the effectiveness of each proposed component, while the case study highlights UniSID’s capability to capture rich and high-level semantic information within the generated SIDs.

The main contributions of this work are as follows:

\begin{itemize}[leftmargin=*]
    \item We identify the limitations of the prevailing two-stage cascading compression paradigm for SID generation, including objective misalignment, semantic degradation, and error accumulation.
    \item We propose UniSID to resolve the above three limitations via three novel designs: end-to-end joint SID-embedding optimization, an advertisement-enhanced input schema, and full-context multi-layer SID prediction, respectively.
    \item We design multi-granularity contrastive learning and a summary-based reconstruction to empower SIDs with different granularities and high-level semantic information.
\end{itemize}
\begin{figure*}[htbp]
\centering
\includegraphics[width=0.95\linewidth]{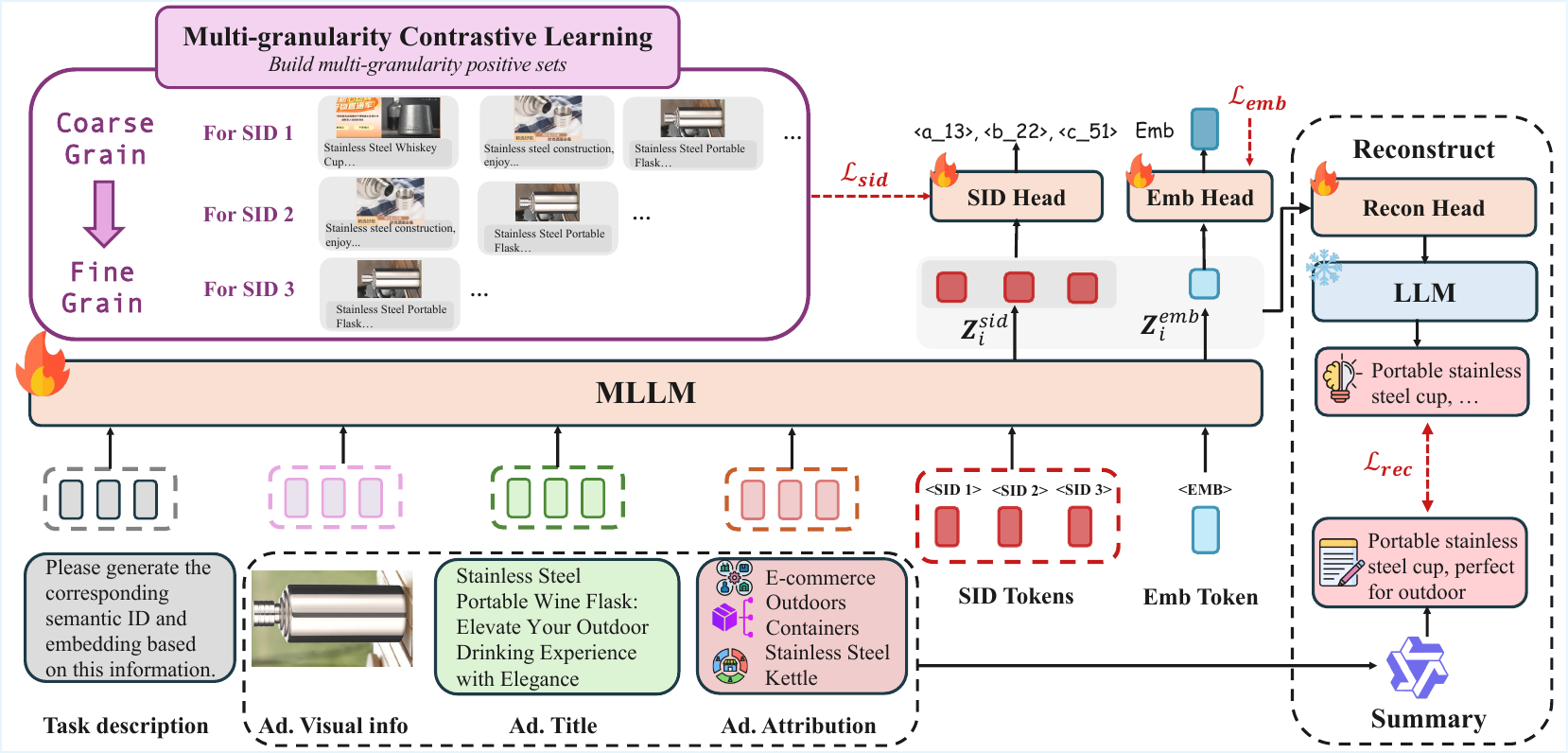} 
\caption{The framework of UniSID}
\label{fig_model}    
\end{figure*}
\section{Preliminaries}

\stitle{Generative Recommendation.}
GR reformulates the recommendation as a sequence generation problem, where the model directly generates items conditioned on user behavior history, rather than ranking items from a candidate set.
Let $u$ denote a user, $c$ denote contextual information, and
$\mathbf{i}_{1:T} = (i_1, i_2, \dots, i_T)$ denote the historical interaction sequence,
where $i_t \in \mathcal{I}$ is an item from the item space $\mathcal{I}$.
GR models the generative process of user behavior autoregressively:
$p_\theta(\mathbf{i}_{1:T} \mid u, c)
=
\prod_{t=1}^{T}
p_\theta(i_t \mid \mathbf{i}_{<t}, u, c),$ where $\mathbf{i}_{<t} = (i_1, \dots, i_{t-1})$.
The model is typically trained using next-token prediction to maximize the likelihood of observed interaction sequences.

\stitle{Semantic ID.}
SID is a discrete token sequence used to represent items in GR, enabling recommendation to be formulated as a sequence generation problem. 
Existing GR methods typically construct SIDs using RQ. 
Formally, for an item $i$ with embedding $\mathbf{Z}_i$, its SID is defined as 
$s_i = \{ s_i^{1}, \dots, s_i^{L} \},$
where $L$ denotes the number of quantization levels. 
At each level $l$, a code is selected from a level-specific codebook 
$\mathcal{C}^{l} = \{ \mathbf{c}_1^{l}, \mathbf{c}_2^{l}, \dots, \mathbf{c}_K^{l} \}.$
RQ constructs SID hierarchically via residual quantization by initializing $\mathbf{r}_i^{1} = \mathbf{Z}_i$ and iteratively selecting
$s_i^{l} = \arg\min_k \left\| \mathbf{r}_i^{l} - \mathbf{c}_k^{l} \right\|_2^2,$
followed by residual update 
$\mathbf{r}_i^{l+1} = \mathbf{r}_i^{l} - \mathbf{c}_{s_i^{l}}^{l}.$
After $L$ levels of quantization, the selected codes form the SID $s_i$.

\section{Methodology}
Figure~\ref{fig_model} provides an overview of UniSID, a unified SID generation framework for ad GR. UniSID consists of three key components: an advertisement-enhanced input schema, a multi-granularity contrastive learning {strategy}, and a summary-based ad reconstruction mechanism. The advertisement-enhanced input schema integrates heterogeneous advertising signals into a unified token sequence, including ad instruction prompts, visual content, textual descriptions, structured ad attributes (e.g., industry and category), as well as a set of learnable SID tokens and embedding tokens. These tokens are jointly processed by a shared multimodal large language model (MLLM), which encodes all inputs into hidden states. Based on the output of the MLLM, UniSID employs two task-specific heads to generate the SID token and item embedding, respectively. The generated SIDs are then optimized through multi-granularity contrastive learning to enforce semantic consistency at different SID granularities. In addition, a summary-based ad reconstruction mechanism further compels the SIDs to capture high-level semantic information. In the following sections, we introduce each component of UniSID in detail.

\subsection{Advertisement-Enhanced Input Schema}
Current GR typically focuses on unstructured modalities such as images, texts, or videos. However, in advertising scenarios, structured ad attributes provide essential semantic constraints that are difficult to infer solely from visual or textual content. Images and texts in advertisements exhibit inherent semantic ambiguity. For example, an advertisement containing an image of a bottle with the text ``natural'' may correspond to a beverage ad, a skincare product, or a health supplement. Without explicit ad attributes, such ambiguity cannot be reliably resolved. In UniSID, we construct a comprehensive multimodal ad feature by integrating visual content, textual descriptions, and structured ad attributes.  

\stitle{{Ad Instruction Prompt.}}
The ad instruction prompt $x_i^{\text{task}}$ is a textual sequence that explicitly specifies the SID and embedding generation task. It guides the model to focus on learning SIDs and embedding rather than generic text generation. In practice, we use a concise instruction such as: \emph{``Given the following advertisement information, please generate the corresponding Semantic IDs and embedding.''}

\stitle{{Images and Text.}}
The image $x_i^{\text{img}}$ represents the visual content of the advertisement, while the text $x_i^{\text{text}}$ typically corresponds to the ad title or description. These unstructured modalities provide rich semantic cues about the appearance and intent of the ad. By jointly encoding visual and textual information, UniSID ensures that multimodal semantic signals are effectively integrated into the SID space.

\stitle{{Ad Attributes.}}
Ad attributes $x_i^{\text{att}}$ consist of a set of structured features that precisely define the advertisement. In particular, we incorporate industry and multi-level category information to reduce semantic ambiguity. For example, for a product such as a water cup, the industry category is \emph{general e-commerce}, while the hierarchical category path is \emph{daily necessities $\rightarrow$ tableware $\rightarrow$ drinkware $\rightarrow$ water cup}. These structured attributes provide explicit semantic constraints that cannot be reliably inferred from images or texts alone.

\stitle{{SID Tokens.}}
Standard MLLMs are primarily designed for textual generation and lack the capability to directly produce discrete SIDs. To bridge this gap, we incorporate multiple learnable SID tokens following the advertising inputs. During the next-token prediction process, these tokens aggregate multimodal and attribute-rich features through the shared MLLM. The resulting representations are then mapped by a specialized SID head to generate discrete SIDs. This design obviates the need for cascading compression, establishing a robust foundation for end-to-end SID generation.

\stitle{{Embedding Token.}}
Positioned after the SID tokens, we further introduce an embedding token. Following the same processing logic as the SIDs, this token is updated by the MLLM and subsequently projected via an embedding head. Crucially, by leveraging the next-token prediction mechanism, the embedding token is conditioned on the preceding SID sequence. This allows it to integrate both raw advertising content and {coarse-to-fine} semantic information, resulting in richer representations than those derived from isolated raw data. Furthermore, this architectural design enables SIDs and embeddings to mutually reinforce each other through joint optimization, significantly boosting the robustness and quality of the embedding.

\subsection{Unified SID and Embedding Generation}
Formally, let $X_i$ denote the concatenated input token sequence of an ad item $i$, including instruction tokens, image tokens, text tokens, ad attribute tokens, SID tokens, and embedding token. The shared MLLM encodes $X_i$ into a sequence of hidden states:
\begin{equation}
\mathbf{Z}_i = \text{MLLM}(X_i),
\end{equation}
where $\mathbf{Z}_i$ represents the contextualized representations of all tokens.

We then extract the representations at the positions of the SID tokens and embedding token, denoted as $\mathbf{Z}_i^{\text{SID}}$ and $\mathbf{Z}_i^{\text{Emb}}$, respectively. These token-specific representations capture aggregated semantic information from the entire accessible context. To generate SIDs and item embeddings, UniSID adopts a dual-head projection design. Specifically, the SID head projects $\mathbf{Z}_i^{\text{SID}}$ into the SID embedding space, while the embedding head projects $\mathbf{Z}_i^{\text{Emb}}$ into the item embedding space,
\begin{equation}
\mathbf{z}_i^{\text{SID}} = f_{\text{SID}}(\mathbf{Z}_i^{\text{SID}}),
\end{equation}
\begin{equation}
\mathbf{z}_i^{\text{Emb}} = f_{\text{Emb}}(\mathbf{Z}_i^{\text{Emb}}),
\end{equation}
where $f_{\text{SID}}(\cdot)$ and $f_{\text{Emb}}(\cdot)$ are lightweight linear projection heads.

For SID generation, each item is associated with a multi-layer SID consisting of $L$ semantic layers. For item $i$, its projected SID representation $\mathbf{z}_i^{\text{SID}}$ is split into $L$ layer-wise SID logits: $\mathbf{z}_i^{\text{SID}} = \{\mathbf{z}_i^{1}, \mathbf{z}_i^{2}, \dots, \mathbf{z}_i^{L}\},$
where $\mathbf{z}_i^{l}$ denotes the SID logits of item $i$ at the $l$-th semantic layer. 
The discrete SID token at layer $l$ is obtained by applying an \texttt{argmax} operation over the corresponding logits:
\begin{equation}
s_i^{l} = \arg\max \big( \mathbf{z}_{i}^{l} \big), \quad l = 1, \dots, L,
\end{equation}
where $s_i^{l}$ denotes the SID token of item $i$ at the $l$-th layer. 
By concatenating the layer-wise tokens $\{s_i^{1}, \dots, s_i^{L}\}$, UniSID directly generates the SID sequence $s_i$ for item $i$.

\subsection{Multi-granularity Contrastive Learning}

{According to the hierarchical nature of SIDs, the selection of positive samples in contrastive learning for SID optimization should also account for the granularity of each semantic level. Specifically, rather than defining a fixed set of positive samples for all layers, we introduce a multi-granularity contrastive learning strategy that adaptively determines the positive relationships at each semantic level according to ad relevance. In particular, as the hierarchy deepens, the required similarity between query and positive samples increases, fully reflecting the hierarchical nature of SIDs.} Therefore, we construct distinct positive sample sets for each SID granularity and apply contrastive learning independently at different SID levels. This design explicitly enforces each SID to capture semantics that are appropriate to its corresponding granularity, preventing fine-grained SIDs from absorbing coarse-level noise and avoiding semantic ambiguity across hierarchical SID tokens.

Specifically, given an ad item $i$ with its SID representation $\mathbf{Z}_i^{\text{sid}}$, 
we perform contrastive learning at multiple SID granularity levels $l \in \{1, \dots, L\}$. 
At each granularity level $l$, we define a positive set $\mathbf{P}_l$ consisting of ad items 
that share the same semantic category with item $i$ at level $l$, and a candidate set $\mathbf{A}_l$ 
that includes both positive and negative samples at the same granularity. 
We optimize the following multi-granularity contrastive objective:
\begin{equation}
\mathcal{L}_{\text{sid}} =
\frac{1}{L} \sum_{l=1}^{L}
\frac{-1}{|\mathbf{P}_l|}
\sum_{p \in \mathbf{P}_l}
\log
\frac{
\exp \left( \mathrm{sim}(\mathbf{z}_i^{l} \cdot \mathbf{z}_p^{l}) / \tau \right)
}{
\sum_{a \in \mathbf{A}_l}
\exp \left( \mathrm{sim}(\mathbf{z}_i^{l} \cdot \mathbf{z}_a^{l}) / \tau \right)
},
\end{equation}
where $\mathrm{sim}(\cdot,\cdot)$ denotes cosine similarity, $\mathbf{z}_i^{l}$ denotes the SID embedding of item $i$ at granularity level $l$, 
$\mathbf{z}_p^{l}$ and $\mathbf{z}_a^{l}$ denote the embeddings of a positive sample $p$ 
and a candidate sample $a$ at the same level, respectively. 
$\mathbf{P}_l$ and $\mathbf{A}_l$ represent the positive set and the candidate set at level $l$, 
and $\tau$ is a temperature hyper-parameter.
Through multi-granularity contrastive supervision, UniSID achieves accurate semantic disentanglement 
across SID levels, enabling coarse-to-fine SIDs to form a consistent and well-structured semantic hierarchy.

The embedding is optimized using a standard contrastive learning objective.
Given a positive pair $(i, j)$ and a set of negative samples $\mathcal{N}_i$, the embedding contrastive loss is defined as:
\begin{equation}
\mathcal{L}_{\mathrm{emb}}
= - \log
\frac{
\exp\left( \mathrm{sim}(\mathbf{z}_i^{\text{Emb}}, \mathbf{z}_j^{\text{Emb}}) / \tau \right)
}{
\sum_{k \in \mathcal{N}_i}
\exp\left( \mathrm{sim}(\mathbf{z}_i^{\text{Emb}}, \mathbf{z}_k^{\text{Emb}}) / \tau \right)
},
\end{equation}
This objective encourages embeddings of semantically similar advertisements to be closer while pushing apart dissimilar ones.

\subsection{Summary-based Ad Reconstruction}

To further enhance the effectiveness of SID in complex advertising scenarios, we propose a summary-based ad reconstruction mechanism. The core motivation is that raw advertising data, even with multimodal content and structured attributes, may not explicitly expose high-level semantic information that is critical for accurate ad understanding. By first summarizing ad attributes into deeper semantic information and then reconstructing them through generated SIDs, we explicitly encourage SIDs to capture latent high-level semantics that are not directly observable in raw data.

\stitle{{Ad Attribute Summary.}}
The summary stage aims to infer latent high-level semantic information from structured ad attributes. Specifically, we leverage industry and hierarchical category information to reason about the ad semantics under the guidance of a task-specific prompt. A frozen LLM is used to summarize the ad attributes into a semantic summary that is not explicitly present in the raw advertising data.

Formally, given the structured ad attributes of item $i$, the summary is generated as:
\begin{equation}
\mathbf{s}_i^{\text{sum}} = \text{LLM}_{\text{sum}}\big( \text{Prompt}_{\text{sum}},\; x_i^{\text{att}}\; \big),
\end{equation}
where $\mathbf{s}_i^{\text{sum}}$ denotes the generated semantic summary for item $i$, and $\text{Prompt}_{\text{sum}}$ provides instruction guidance for semantic summarization. Detailed $\text{Prompt}_{\text{sum}}$ is provided in Figure \ref{fig:prompt}.

\stitle{{Summary Reconstruction.}}
Given the generated semantic summary $\mathbf{s}_i^{\text{sum}}$, UniSID reconstructs it using the learned SIDs $\mathbf{Z}_i^{\text{SID}}$ and item embedding $\mathbf{Z}_i^{\text{Emb}}$. Specifically, we first concatenate the SID embedding and the item embedding to form a unified representation:
\begin{equation}
\mathbf{Z}_i = \big[ \mathbf{Z}_i^{\text{SID}} \, ; \, \mathbf{Z}_i^{\text{Emb}} \big],
\end{equation}
where $[\cdot;\cdot]$ denotes vector concatenation. The combined representation is then projected through a reconstruction head to obtain a hidden state:
\begin{equation}
\mathbf{h}_i^{\text{rec}} = f_{\text{recon}}(\mathbf{Z}_i),
\end{equation}
where $f_{\text{rec}}(\cdot)$ is a lightweight reconstruction projection head.

The hidden state $\mathbf{h}_i^{\text{rec}}$ is subsequently used as the conditioning input to an LLM, which reconstructs the semantic summary under a next-token prediction paradigm. The reconstruction objective is optimized via a standard cross-entropy loss:
\begin{equation}
\mathcal{L}_{\text{rec}} = - \sum_{t=1}^{|\mathbf{s}_i^{\text{sum}}|}
\log p \big( s_{i,t}^{\text{sum}} \mid \mathbf{h}_i^{\text{rec}},\; s_{i,<t}^{\text{sum}} \big),
\end{equation}
where $s_{i,t}^{\text{sum}}$ denotes the $t$-th token of the summary sequence.

By reconstructing high-level semantic summaries solely from SIDs and embeddings, this paradigm explicitly encourages SIDs to encode discriminative and high-level semantic information that is not directly available in raw advertising data, thereby improving their effectiveness in complex advertising scenarios.


\subsection{Joint Optimization}

Unlike prior two-stage paradigms that generate SIDs via cascaded embedding quantization, our method jointly learns semantic IDs and embeddings in an end-to-end manner.
This unified design allows SIDs to be directly induced from raw advertisement data, effectively avoiding objective inconsistency and semantic information loss caused by two-stage compression.

Specifically, we jointly optimize three complementary objectives: 
(i) a multi-granularity contrastive loss for SIDs, 
(ii) a contrastive loss for embeddings, and 
(iii) a reconstruction loss derived from the summary-based ad reconstruction mechanism.
The overall training objective is formulated as:
\begin{equation}
\mathcal{L}_{\mathrm{total}}
= \mathcal{L}_{\mathrm{sid}}
+ \mathcal{L}_{\mathrm{emb}}
+ \lambda \mathcal{L}_{\mathrm{rec}},
\end{equation}
{where $\lambda$ is the hyperparameter of reconstruction loss.} The analysis of $\lambda$ is provided in the Appendix \ref{app_sen}. Additional discussions on efficiency and limitations are in Appendix \ref{discussion}.

\section{Experiments}
In this section, we conduct extensive experiments to validate the superiority of our model UniSID in various real world datasets.
\subsection{Experimental Setup}
\stitle{{Datasets.}}
We evaluate UniSID on both industrial-scale advertising datasets and a widely-used public dataset. Detailed statistics of all datasets are summarized in Table~\ref{tab:ad_data}.
Specifically, we first conduct experiments on two real-world industrial advertising datasets, {Ad-60W} and {Ad-100W}, collected from large-scale ad recommendation systems of Tencent. 
These datasets contain rich multimodal signals and complex hierarchical category structures, making them well-suited for validating the effectiveness of SID generation in realistic and challenging advertising scenarios. We will release the datasets to promote reproducibility and future research in the community. More data samples are shown in Appendix \ref{app_data}.
To further assess the generality of UniSID, we additionally adopt a public dataset that is commonly used in recommendation research \cite{wang2024letteer,sun2019bert4rec}.
{Beauty} is a subset of the Amazon Review dataset \cite{ni2019justifying}, which contains user--item interaction histories on beauty products.
Following standard practice, we apply commonly adopted preprocessing strategies for this public dataset. 

\begin{table}[t]
\caption{Characteristics of two real-world industrial advertising datasets and Beauty dataset.}
\label{tab:ad_data}
\scalebox{0.95}{
\begin{tabular}{@{}l|llll@{}}
\toprule
\textbf{Dataset} & \textbf{Train}   & \textbf{Test}   & \textbf{Modal}                                                               & \textbf{Task}                                                                                                                   \\ \midrule
Ad-60W  & 500,000 & 100,000 & \begin{tabular}[c]{@{}l@{}}Text, Image\\ Ad\_att\end{tabular}        & \begin{tabular}[c]{@{}l@{}}SID Quality\\ Ad-Retrival\end{tabular} \\
Ad-100W  & 1,000,000 & 1,000,000 & \begin{tabular}[c]{@{}l@{}}Text, Image\\ Video,  Ad\_att\end{tabular} & Next-Ad-Pre \\ 
Beauty  & 12,101 & 12,101 & \begin{tabular}[c]{@{}l@{}}Text, Image\\ \end{tabular} & Next-Item-Pre \\
\bottomrule
\end{tabular}}
\end{table}

\stitle{{Baselines.}} We compare UniSID with representative baselines from three categories.
\begin{itemize}[leftmargin=*]
    \item To evaluate the effectiveness of SID generation, we adopt state-of-the-art SID construction methods based on residual quantization, including {RQ-VAE} \cite{rajput2023recommender} and {RQ-KMeans} \cite{luo2025qarm}.
These methods follow the embedding-then-SID two-stage paradigm and are widely used in existing GR frameworks.
\item To assess the quality of embeddings produced by UniSID, we compare against advanced multi-modal embedding methods, including {GME} \cite{zhang2024gme}, LamRA \cite{liu2025lamra}, and {VLM2Vec2} \cite{meng2025vlm2vec}.
These approaches focus on learning unified representations from multi-modal ad content.
\item Finally, we include representative recommendation baselines on the Beauty dataset, covering both discriminative and generative paradigms.
Specifically, we consider classical DLRMs, including Bert4Rec \cite{sun2019bert4rec}, LightGCN \cite{he2020lightgcn}, and SASRec \cite{kang2018sasrec}, as well as GR methods, including BIGRec \cite{bao2025bigrec}, P5-SemID \cite{hua2023P5}, TIGER \cite{rajput2023recommender} and  LETTER \cite{wang2024letteer}.
This comprehensive set of baselines enables a fair and thorough evaluation of UniSID from multiple perspectives.
\end{itemize}

\stitle{{Evaluation setting.}}
We evaluate UniSID from three complementary perspectives. 
\begin{itemize}[leftmargin=*]
    \item SID Evaluation. We conduct SID evaluation on two real-world advertising datasets, Ad-60W and Ad-100W, focusing on both SID quality and SID-based ad recommendation performance. For SID quality, we adopt \textit{V-measure} to assess the clustering quality of SIDs across three hierarchical levels, where the finest-grained category labels are used as ground truth. To evaluate SID performance, we use the next-ad prediction task as the downstream evaluation, measuring the effectiveness of generated SIDs with the Hit Rate (HR@K) metric.

    \item {Embedding Evaluation.}
To evaluate the quality of the generated embeddings, we follow the evaluation protocol of VLM2Vec2 \cite{meng2025vlm2vec}.
Specifically, we reformulate the advertising dataset into a ranking retrieval task, where each query is paired with one positive target item and 999 sampled negative items. Embedding quality is measured using Recall (R@K).
\item {Public Dataset Evaluation.}
For the Beauty dataset, we adopt standard evaluation protocols commonly used in recommendation research. 
Recommendation performance is measured using R@K and NDCG@K (N@K).
\end{itemize}

\subsection{Overall Performance}

\begin{table}[tbp]
\centering
\caption{Performance comparison of SID quality (V-Measure) and next-ad prediction (HR@K) between RQ-based methods and UniSID on the Ad-60W and Ad-100W datasets.
The best is bolded, and the second-best is underlined. The bottom row is the relative improvement of UniSID over the best baseline.}
\label{tab:sid}
\scalebox{0.85}{
\begin{tabular}{@{}c|ccccc@{}}
\toprule
\textbf{Task} & \textbf{Model} & \textbf{RQ-VAE} & \textbf{RQ-Kmeans} & \textbf{UniSID} & \textbf{Imp (\%)} \\ \midrule
\multirow{3}{*}{\begin{tabular}[c]{@{}c@{}}SID \\ Quality\end{tabular}} & layer1 & 0.6769 & \underline{0.6887} & \textbf{0.7015} & 1.86 \\
 & layer2 & 0.6908 & \underline{0.6918} & \textbf{0.7132} & 3.09 \\
 & layer3 & 0.6863 & \underline{0.6955} & \textbf{0.7045} & 1.29 \\ \midrule
\multirow{4}{*}{\begin{tabular}[c]{@{}c@{}}SID \\ Performance\end{tabular}} & HR@1 & 0.0416 & \underline{0.0434} & \textbf{0.0449} & 3.46 \\
 & HR@5 & 0.0735 & \underline{0.0758} & \textbf{0.0793} & 4.62 \\
 & HR@10 & 0.1077 & \underline{0.1122} & \textbf{0.1167} & 4.01 \\
 & HR@20 & 0.1675 & \underline{0.1754} & \textbf{0.1813} & 3.36 \\ \bottomrule
\end{tabular}}
\end{table}

\stitle{{SID Quality Evaluations.}}
As summarized in Table \ref{tab:sid}, UniSID consistently outperforms state-of-the-art RQ-based methods in SID quality across all three semantic granularities (layer1–layer3) on the Ad-60W industrial dataset. These results validate the efficacy of the UniSID framework in learning SIDs directly from raw advertising data via an end-to-end approach. Specifically, UniSID achieves v-measure improvements of 1.86\%, 3.09\%, and 0.90\% over the strongest baseline, RQ-Kmeans, and surpasses RQ-VAE by 3.63\%, 3.24\%, and 2.65\% across the respective layers.
The performance gains suggest that UniSID generates more consistent and structurally robust SID representations across varying levels of semantic abstraction. This superiority is primarily attributed to its unified end-to-end modeling, which mitigates the inherent information loss found in two-stage paradigms caused by objective misalignment and cascading compression. Consequently, UniSID facilitates a more effective integration of advertising semantics into the SID space, significantly enhancing both the homogeneity and semantic integrity of the SID.

\begin{table}[tbp]
\centering
\caption{Overall performance comparison of embedding quality between MM embedding methods and UniSID on the Ad-60W, evaluated by R@K. The best is bolded, and the second-best is underlined. The bottom row is the relative improvement of UniSID over the best baseline.}
\label{tab:embedding}
\scalebox{0.9}{
\begin{tabular}{@{}lcccc@{}}
\toprule
\multicolumn{1}{l}{\textbf{Model}}                 & \textbf{R@1} & \textbf{R@5} & \textbf{R@10} & \textbf{R@20} \\ \midrule
\multicolumn{1}{l|} {\textbf{GME-Qwen2-VL-2B}} & 0.2396        & 0.3195        & 0.4073         & 0.4651         \\
\multicolumn{1}{l|}{\textbf{GME-Qwen2-VL-7B}} & 0.2545              & 0.3620              & 0.4312               & 0.4830               \\
\multicolumn{1}{l|}{\textbf{LamRA-Qwen2-VL-7B}}   & 0.2896              &0.4051               & 0.4592               & 0.5171 \\
\multicolumn{1}{l|}{\textbf{LamRA-Qwen2.5-VL-7B}}     & 0.3112              &0.4280               & 0.4745               & 0.5390               \\
\multicolumn{1}{l|}{\textbf{VLM2Vec2-Qwen2-VL-2B}}   & 0.2692        & 0.3810        & 0.4461         & 0.5094         \\ 
\multicolumn{1}{l|}{\textbf{VLM2Vec2-Qwen2.5-VL-3B}} & \underline{0.3238}
& \underline{0.4475}        & \underline{0.5062}         & \underline{0.5597}         \\
\midrule
\multicolumn{1}{l|}{\textbf{UniSID-Qwen2.5-VL-3B}}          &      \textbf{0.4710}         &     \textbf{0.5735}          &    \textbf{0.6102}            &   \textbf{0.6488}             \\ \midrule
\multicolumn{1}{l|}{\textbf{Imp (\%)}}          &      {45.46}         &     {28.16}          &    {20.55}            &   {15.92} \\
\bottomrule
\end{tabular}}
\end{table}

\stitle{{SID Performance Evaluations.}}
As illustrated in Table \ref{tab:sid}, we further evaluate the performance of various SID generation methods on the industrial advertising dataset Ad-100W, employing a next-ad prediction task. The overall results demonstrate that UniSID consistently outperforms all baseline methods across all HR@K metrics, exhibiting superior SID representation capabilities in real ad scenarios. Specifically, compared to RQ-Kmeans, UniSID achieves performance gains of 3.46\%, 4.62\%, 4.01\%, and 3.36\% in terms of HR@1, HR@5, HR@10, and HR@20, respectively. Against RQ-VAE, it yields improvements of 7.93\%, 7.89\%, 8.36\%, and 8.24\%. 

This performance edge is primarily attributable to two factors: First, UniSID adopts a unified end-to-end modeling approach to generate SIDs directly from raw advertising data, effectively mitigating the semantic degradation inherent in two-stage cascading compression. Second, the integration of a multi-granularity contrastive learning strategy and a summary-based ad reconstruction mechanism further strengthens the semantic expressiveness of the SIDs. While the former ensures precise and consistent discriminative power across varying semantic granularities, the latter enables SIDs to capture latent high-level semantic features not explicitly present in the raw data. Through the synergy of these mechanisms, UniSID produces richer and more precise SID representations in complex and heterogeneous advertising environments, thereby significantly boosting overall recommendation performance.

\stitle{{Embedding Quality Evaluations.}}
Table \ref{tab:embedding} presents the evaluation results of embedding retrieval performance on the Ad-60W industrial dataset. The results indicate that UniSID significantly outperforms existing multi-modal embedding methods across all Recall@K metrics, demonstrating superior semantic representation capabilities. Specifically, compared to VLM2Vec2, UniSID achieves performance gains of 45.46\%, 28.16\%, 20.55\%, and 5.92\% in terms of Recall@1, Recall@5, Recall@10, and Recall@20, respectively. This performance boost is primarily driven by UniSID’s unified end-to-end architecture. On one hand, the joint generation of SIDs and embeddings fosters a synergistic optimization during training; this allows the embeddings to explicitly incorporate the hierarchical semantic information encoded within the SIDs, leading to more discriminative representations. On the other hand, the embedding generation process not only models the multi-modal information from raw advertising data but also observes the SID structure from coarse to fine granularities. This process can be viewed as an implicit Chain-of-Thought (CoT) semantic guidance, which progressively facilitates the extraction of more granular semantic features.

\begin{table}[tbp]
\caption{Comparison between UniSID and baselines on the Beauty dataset in terms of R@K and N@K. The best is bolded, and the second-best is underlined. The bottom row is the relative improvement of UniSID over the best baseline.}
\centering
\setlength{\tabcolsep}{3mm}{ 
\scalebox{0.9}{
\begin{tabular}{l|cccc} 
\toprule
\multicolumn{1}{l|}{\textbf{Model}} & 
\textbf{R@5} & \textbf{R@10} & \textbf{N@5} & \textbf{N@10} \\ \midrule
\textbf{Bert4Rec} & 0.0203 & 0.0347 & 0.0124 & 0.0170 \\
\textbf{LightGCN} & 0.0305 & 0.0511 & 0.0194 & 0.0260 \\
\textbf{SASRec}   & 0.0380 & 0.0588 & 0.0246 & 0.0313 \\
\textbf{BigRec}   & 0.0243 & 0.0299 & 0.0181 & 0.0198 \\
\textbf{P5-SemID}       & 0.0393 & 0.0584 & 0.0273 & 0.0335 \\
\textbf{TIGER-LETTER}    & \underline{0.0431} & \underline{0.0672} & \underline{0.0286} & \underline{0.0364} \\\midrule
\textbf{TIGER}    & 0.0395 & 0.0610 & 0.0262 & 0.0331 \\

\textbf{TIGER-UniSID} & \textbf{0.0482} & \textbf{0.0741} & \textbf{0.0323} & \textbf{0.0406} \\ \midrule
\textbf{\textbf{Imp (\%)}} & {11.83} & {10.27} & {12.94} & {11.54} \\ 
\bottomrule
\end{tabular}}
}
\label{tab:BeautyOnly}
\end{table}

\stitle{{Public Dataset Results.}}
Table \ref{tab:BeautyOnly} reports the experimental results \footnote{Baseline results are from LETTER} of UniSID in public recommendation benchmarks: Amazon Beauty. To ensure a fair comparison and isolate the impact of SIDs, we integrated UniSID into the Tiger framework by replacing its original RQ-VAE SID generation module while maintaining all other training and evaluation protocols.
The results demonstrate that UniSID-generated SIDs consistently achieve SOTA performance, outperforming both classical DLRM-based models and GR baselines. Specifically, compared to the original Tiger framework, UniSID yields significant improvements of 22.03\%, 21.48\%, 23.28\%, and 22.66\% across Recall@1, 5, 10, and 20, respectively. Relative to Tiger-LETTER, which replaces Tiger's SIDs with LETTER, UniSID still yields notable gains of 11.83\%, 10.27\%, 12.94\%, 11.54\%, and 11.65\% across the same metrics. These findings underscore that UniSID not only generates high-quality SIDs for complex industrial advertising scenarios but also seamlessly generalizes to standard public benchmarks. This validates the robust scalability and strong generalization capabilities of UniSID across diverse data distributions and application contexts.

\begin{figure}[tbp]
\centering
\includegraphics[width=0.95\columnwidth]{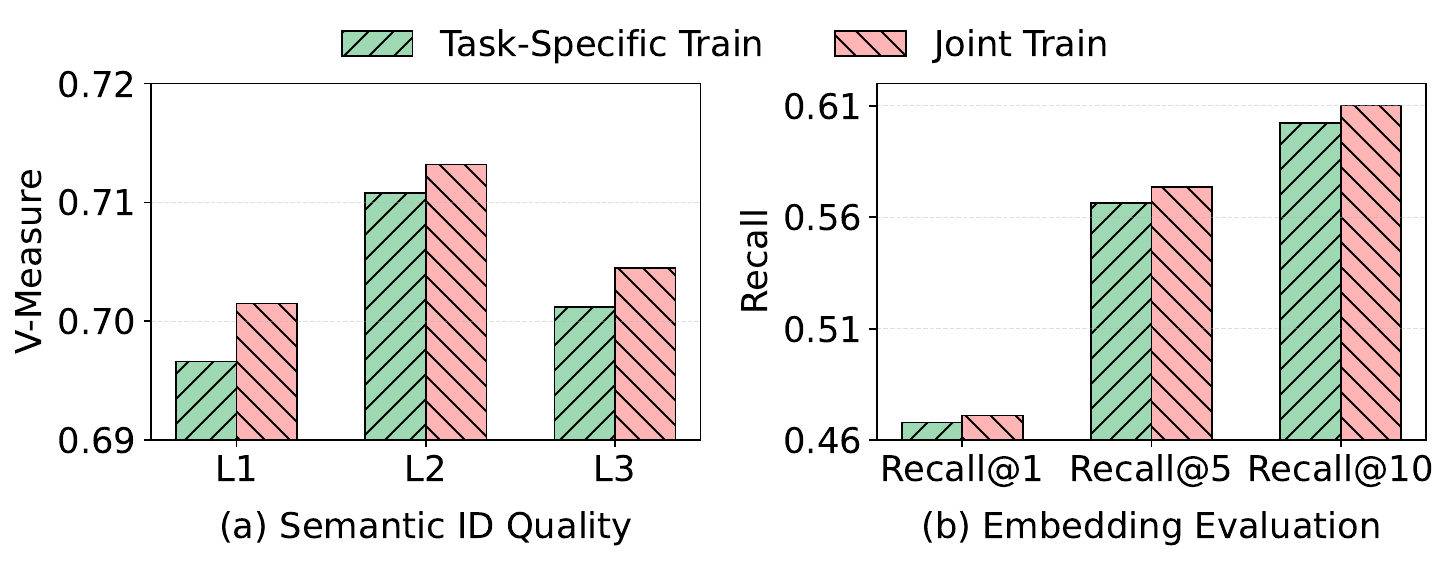}
\caption{Comparison between joint training and task-specific separate training on the Ad-60W data}
\label{fig:joint_train} 
\end{figure}

\subsection{Ablation Study}
We conduct ablation studies on UniSID to systematically evaluate the effectiveness of each proposed component.

\stitle{Effect of Joint Train.}
 Figure \ref{fig:joint_train} presents a comparative analysis between the proposed joint training framework and single-task end-to-end training variants, evaluated across both SID quality and embedding quality metrics. The results indicate that compared to the full UniSID model, variants optimized for a single objective suffer varying degrees of performance degradation across all indicators. These findings highlight the critical role of the joint training mechanism. By simultaneously optimizing SIDs and embeddings, the two components benefit from a reciprocal reinforcement process. Embedding learning is regularized by the hierarchical semantic structure of the SIDs, leading to more structured representations. SID generation leverages richer semantic signals captured within the embeddings. This collaborative optimization effectively enhances the expressive power of both SIDs and embeddings, validating the necessity and efficacy of the joint training architecture.

\stitle{Effect of Contrastive Learning.}
The upper section of Table \ref{tab:abl_loss} compares the performance of various contrastive loss functions, including KL loss, JS loss, BCE loss, and our proposed Multi-Granularity Contrastive Loss (MG Loss) on SID quality assessment. The baseline employs a standard InfoNCE loss as the contrastive objective. Observation reveals significant performance variations across the three SID layers among the different loss functions. While distribution-matching losses like KL and JS can model inter-sample similarity to an extent, they struggle to explicitly distinguish between different semantic granularities in multi-level SID scenarios, leading to suboptimal performance. Although BCE loss provides training stability, its capacity to model hierarchical semantic structures is restricted, resulting in marginal improvements. In contrast, MG Loss achieves superior performance across all hierarchy levels. This excellence stems from its ability to construct granularity-aware positive samples: sharing coarse-grained semantics are treated as positive pairs, whereas they are further differentiated as negative pairs at finer-grained levels. By explicitly modeling this hierarchical consistency, MG Loss compels the SID to capture more precise and accurate semantic structures at each level, significantly enhancing overall SID quality.

\begin{table}[tbp]
\centering
\caption{Ablation study of UniSID's contrastive loss functions and reconstruction designs on the AD-60w dataset.}
\label{tab:abl_loss}
\scalebox{0.9}{
\begin{tabular}{lccl}
\toprule
\textbf{Model}                       & \textbf{L1} & \textbf{L2} & \textbf{L3} \\ \midrule
Baseline (Qwen2.5-VL + InfoNCE Loss) &      0.5889       &      0.6913        &  0.6966  \\ \midrule
\textbf{Contrastive Learning}        &             &             &    \\
\quad + KL Loss                            &   0.5521     &  0.6658     &  0.6906  \\
\quad + JS Loss                            & 0.6030       & 0.6922     &  0.6954  \\
\quad + BCE Loss                            &    0.5951    &   0.6908    &  0.6963  \\
\quad + MG Loss (Ours)                            & \textbf{0.6838}       & \textbf{0.6978}      &  \textbf{0.6967}  \\ \midrule
\textbf{Reconstruction}              &             &             &    \\
\quad + Attributions                        &   0.6999     &  0.7130   & 0.7031   \\
\quad + LLM summary (Ours)                        & \textbf{0.7015}       &   \textbf{0.7132}    & \textbf{0.7045}   \\ \bottomrule
\end{tabular}
}
\end{table}

\stitle{Effect of Reconstruction}
The lower section of Table \ref{tab:abl_loss} investigates the impact of various reconstruction strategies on SID quality. We first evaluate an attribute-augmented reconstruction strategy, which leverages industrial and multi-level category metadata to supervise the reconstruction process via SIDs and embeddings. Experimental results demonstrate significant gains across all three SID levels, indicating that attribute augmentation effectively encourages the SIDs to encode richer, more structured semantic information. Building upon this, we introduce the summary-based ad reconstruction mechanism. This approach first summarizes latent semantic information that is not explicitly present in the raw attribute data; these summaries are then reconstructed using the SIDs and embeddings. This design yields further performance improvements across all levels, validating its efficacy. These results suggest that the summary-based ad reconstruction guides SIDs to capture implicit, high-level semantic features, thereby significantly enhancing their expressive power and semantic integrity within complex advertising contexts.

\begin{table*}[htbp]
\centering

\caption{Case study on the Ad-60W dataset. Attributes information extracted from raw advertising data is highlighted in bold, while latent high-level semantics derived via summary-and-reconstruction are marked in red.}
\renewcommand{\arraystretch}{1.3} 

\newcolumntype{L}{>{\raggedright\arraybackslash}m{0.31\textwidth}} 
\scalebox{0.8}{
\begin{tabularx}{\textwidth}{L|L|L}
\toprule
\rowcolor{gray!10} \centering \textbf{Advertisement Attributes} & \centering \textbf{Summary Results} & \centering \textbf{Reconstruction Results} \tabularnewline \midrule

\centering
\vspace{5pt}
\includegraphics[width=0.8\linewidth]{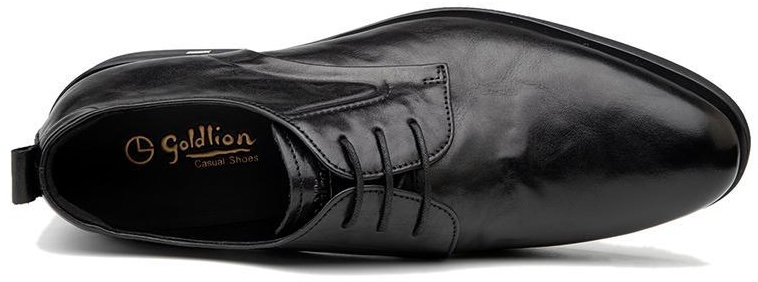} \par
\vspace{5pt}
\begin{flushleft}
\small
{Title:} Goldlion Men's Shoes, Genuine Leather Lace-up, Stylish Business Pointed-toe Shoes \\
{Industry:} General E-commerce \\
{Level 1:} Apparel \& Accessories \\
{Level 2:} Trendy Men's Shoes \\
{Level 3:} Low-top Shoes \\
\end{flushleft}
& 
This advertisement sells Goldlion men's shoes. The industry is \textbf{general e-commerce}, and the primary category is \textbf{apparel and accessories}. It showcases \textbf{a pair of black genuine leather, lace-up, pointed-toe formal business shoes}, emphasizing a \textbf{stylish business style} for \textcolor{red}{\textbf{male consumers}} who \textcolor{red}{\textbf{pursue quality and a professional image}}.
& 
This advertisement sells men's shoes. The industry is \textbf{general e-commerce}, and the primary category is \textbf{apparel and accessories}. The ad showcases a pair of {\textbf{black leather men's business shoes}}, emphasizing their \textcolor{red}{\textbf{comfort, fashion, and versatility}} for \textcolor{red}{\textbf{male consumers}} who \textcolor{red}{\textbf{pursue a quality lifestyle}}. \tabularnewline \bottomrule

\end{tabularx}}
\label{table:case}
\end{table*}

\subsection{Case Study}
To qualitatively demonstrate the effectiveness of UniSID in capturing rich semantic information, we conduct a case study by examining reconstruction results induced by the generated SIDs, as shown in Table \ref{table:case}.
Given an advertisement with multimodal inputs and structured attributes, the summary module infers implicit high-level semantics that are not explicitly present in the raw data. Specifically, it identifies the target audience as “male consumers who pursue quality and a professional image,” which goes beyond literal descriptions in the product title and visual content. 
Conditioned on the generated SIDs and embeddings, the reconstruction module successfully recovers both explicit and implicit semantics. Besides preserving core attribute information (e.g., product type and category), it predicts higher-level semantic concepts such as “comfort, fashion, and versatility” and “male consumers who pursue a quality lifestyle,” which align closely with the inferred summary semantics despite not appearing in the original inputs.
Overall, this case study highlights two key properties of UniSID: (1) generated SIDs effectively encode structured advertising attributes and category semantics; and (2) the summary-based reconstruction mechanism encourages SIDs to capture implicit high-level semantics beyond raw data, validating the effectiveness of UniSID for ID generation in complex advertising scenarios. More case studies are provided in the Table \ref{table:cases}.

\section{Related Work}

\stitle{Generative Recommendation}
In recent years, with the success of LLMs in sequence modeling and generation tasks, research on recommendation systems has gradually shifted from a discriminative modeling approach to a generative paradigm \cite{li2024survey,wang2025generative}. A line of work focuses on designing specialized generative models by varying Transformer-style architectures as the backbone and feature construction paradigms to scale up recommendation capacity \cite{zhai2024actions,han2025mtgr,chai2025longer,zhang2025onetrans,huang2025genrank}. In parallel, another stream of research LLMs to empower existing recommender systems by offline generation of high-level semantic features or auxiliary signals, enabling a progressive upgrade of traditional pipelines \cite{chen2024hllm,yan2025lum,yi2025recgpt}, where LLMs serve as external knowledge or feature generators rather than fully replacing the recommendation stack. 
Despite their effectiveness, many of these methods still retain original DLRM-style features or rely on multi-stage cascading paradigms, inheriting issues such as objective misalignment and information bottlenecks \cite{yan2025lum}. More recently, research has shifted toward more unified end-to-end recommendation frameworks, which aim to employ a single model to jointly perform user understanding and recommendation generation by formulating the entire process as next-token prediction \cite{zhou2025onerec,zhou2025onerec2,zhang2025gpr}. These methods exemplify this trend and demonstrate the potential of generative modeling to replace traditional retrieval–ranking pipelines. Despite their generative nature, most existing methods still depend on external SIDs and two-stage paradigms, failing to achieve true end-to-end learning from data.

\stitle{Semantic ID for Recommendation Systems}
SIDs provide a compact discrete representation for large item spaces by mapping items to semantic token sequences, enabling efficient indexing, retrieval, and generation in large-scale recommendation systems \cite{hou2023learning,singh2024better,zheng2024adapting,li2025survey}. Early SID-based recommendation models were primarily retrieval-based. These methods typically construct SIDs through clustering or hashing over item representations and use them as indexing units in retrieval systems \cite{petrov2024recjpq,hou2023learning}. Recommendation is performed by matching user representations with SID-based indices, often in a nearest-neighbor. While effective in improving retrieval efficiency, these approaches treat SID construction as a preprocessing step and do not tightly integrate it with downstream recommendation. With the rise of generative models and insights from scaling laws in LLM, SIDs have increasingly been used as generation targets \cite{zhou2025onerec,zhang2025gpr}.  Generative SID-based methods demonstrate improved flexibility and expressiveness compared to retrieval-based designs, and they form the foundation of many recent GR \cite{li2025survey}. More recent work has predominantly adopted RQ to construct SIDs, including RQ-VAE \cite{rajput2023recommender}, RQ-KMeans \cite{luo2025qarm}, and RQ-KMeans+ \cite{zhang2025gpr}. These methods first learn item embeddings, and then discretize them into multi-level SIDs via residual vector quantization. By progressively quantizing residuals, RQ-based methods can represent coarse-to-fine semantic information in a structured manner. However, these approaches inherently follow a two-stage paradigm, where item embeddings are first learned and then discretized into SIDs. Such a decoupled design prevents SID construction from being jointly optimized with the recommendation objective. 

\section{Conclusion}
In this paper, we present UniSID, a novel framework that unifies SID generation through end-to-end optimization, thereby overcoming the inherent limitations of the prevailing two-stage cascading compression paradigm. By jointly optimizing embeddings and SIDs, our approach ensures that the generated SIDs capture rich and robust semantic information. To further enhance the fidelity of the SID, we incorporate a multi-granularity contrastive learning strategy alongside a summary-based ad reconstruction mechanism. These components empower SIDs to encapsulate both fine-grained and latent high-level semantics. Extensive experiments conducted on two large-scale industrial advertising datasets and a public benchmark demonstrate that UniSID consistently outperforms state-of-the-art baselines. Furthermore, our ablation studies confirm the necessity and effectiveness of each architectural component, while the case study qualitatively validates that UniSID successfully learns authentic and interpretable semantics tailored to real-world advertising contexts.

\bibliographystyle{ACM-Reference-Format}
\bibliography{sample-base}

\appendix

\section{Experiments}

\subsection{Datasets}\label{app_data}
\begin{figure*}[tbp]
\centering
\includegraphics[width=0.9\linewidth]{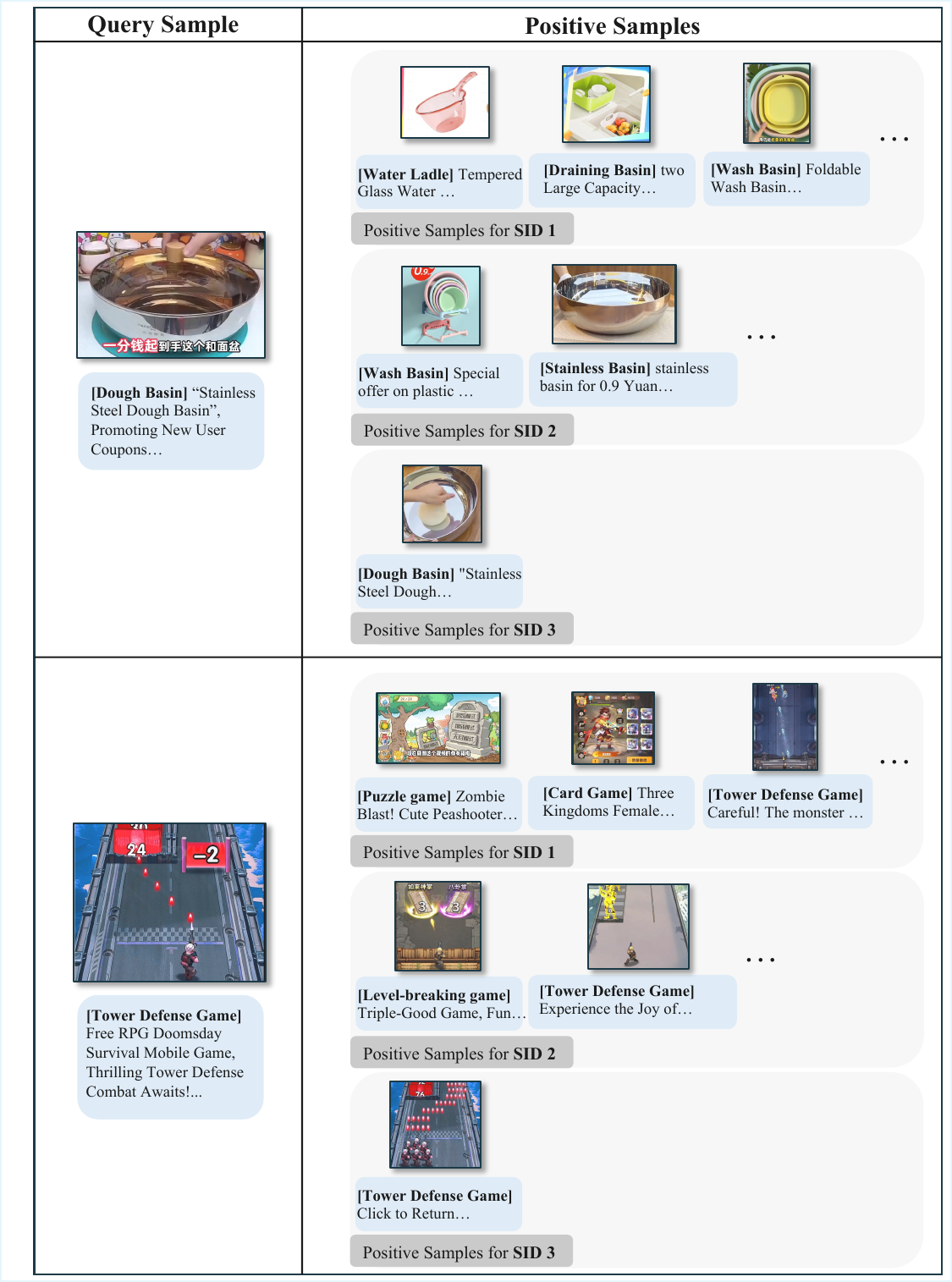}
\caption{Samples for multi-granularity contrastive learning}
\label{fig:dataset_sample}
\end{figure*}

We provide a detailed illustration of the training data construction strategy used for multi-granularity contrastive learning, as shown in Figure \ref{fig:dataset_sample}. 
For each advertisement item, we construct positive samples according to different SID hierarchy levels, enabling the model to learn semantic consistency at multiple granularities.
Specifically, consider the query item \textit{Dough Basin}. 
At the coarse-grained level (SID$_1$), semantically related items such as 
\textit{Water Ladle}, \textit{Draining Basin}, and \textit{Wash Basin} 
are treated as positive samples, since they share similar high-level category semantics. 
At the intermediate level (SID$_2$), more semantically aligned items, including 
\textit{Wash Basin} and \textit{Stainless Basin}, 
are selected as positive samples. 
At the fine-grained level (SID$_3$), the positive sample corresponds to the most specific semantic match, namely \textit{Dough Basin}.

\subsection{Hyperparameter Analysis}\label{app_sen}

\begin{figure}[tbp]
\centering
\includegraphics[width=0.95\columnwidth]{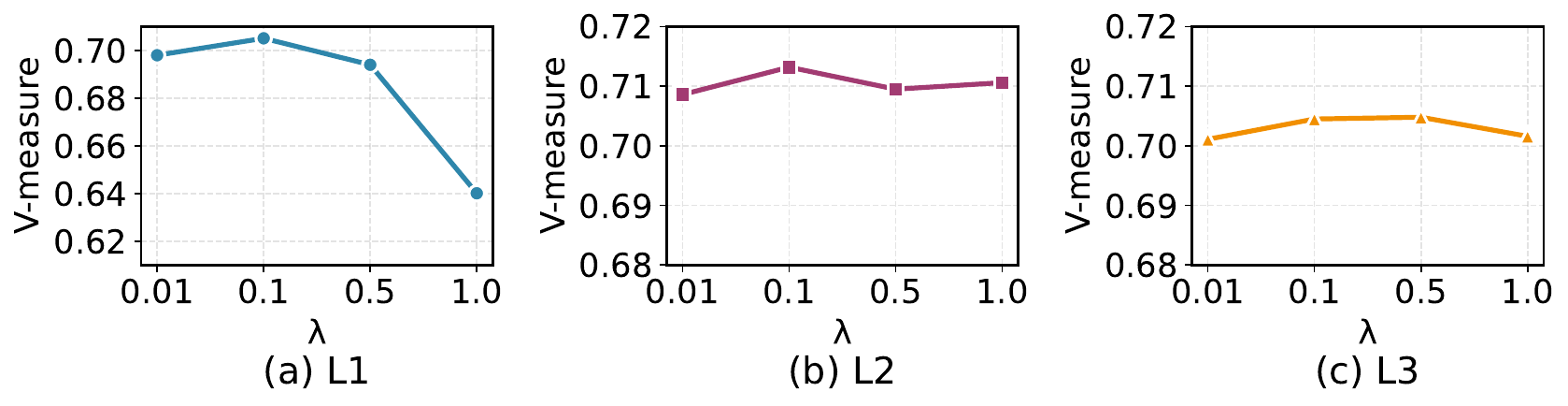}
\caption{The impact of hyperparameter $\lambda$ on SID quality in the Ad-60W dataset.}
\label{fig:par_sen}
\label{tab:joint_train}    
\end{figure}

In this section, we analyze the impact of the reconstruction loss weight $\lambda$ on SID quality.
Specifically, we vary $\lambda$ in $\{0.01, 0.1, 0.5, 1.0\}$ and report the SID quality at three semantic levels (L1, L2, and L3). The results are shown in Table~\ref{fig:par_sen}.
From the table, we observe that the three SID levels exhibit a consistent trend with respect to $\lambda$.
When $\lambda$ is too small, the reconstruction objective provides limited semantic guidance, making UniSID less effective at capturing high-level semantic information, which leads to suboptimal SID quality.
In contrast, when $\lambda$ becomes too large, the reconstruction loss overly interferes with the multi-granularity contrastive learning objective, resulting in degraded SID representations.
When $\lambda$ is set to a moderate value (e.g., $0.1$), the reconstruction loss serves as an effective auxiliary objective that complements contrastive learning and encourages the model to capture richer ad semantic information.
This balance leads to the best overall SID quality across different semantic levels.

\subsection{{Implementation details.}} \label{app_details}
We employ Qwen2.5-VL-3B as our MLLM backbone. The vision encoder (ViT) is frozen, and we perform supervised fine-tuning on the MLLM part. For the reconstruction module, we adopt Qwen2.5-3B as the base LLM with its parameters frozen, optimizing only the reconstruction head. The learning rate is set to 4e-5 with a batch size of 512. In addition, the SID comprises three layers ($L$=3), each with a codebook size of 2,048. The hyperparameter of reconstruction loss $\lambda$ is set to 0.1. All baselines employ the same setting for a fair comparison. 

\section{More Discussion} \label{discussion}
\subsection{Efficiency Analysis} 
We analyze the training efficiency of UniSID in comparison with traditional two-stage RQ-based SID generation methods. 
In two-stage approaches, item embeddings are first learned and then discretized into SIDs through a separate quantization stage. 
In contrast, UniSID performs end-to-end SID generation within a unified framework, while maintaining a comparable computational scale.
The only additional component introduced in UniSID is the LLM used in the reconstruction module. 
Notably, this LLM remains frozen during training and does not participate in parameter updates. 
As a result, it does not introduce additional optimization overhead or training instability. 
The primary trainable components in UniSID remain the MLLM, SID head, embedding head, and reconstruction head.
Therefore, the overall training complexity of UniSID is comparable to RQ-based methods, while providing improved semantic modeling capability through unified SID generation.
\subsection{Limitation}
Despite its effectiveness, UniSID still has several limitations. 
First, the summary-then-reconstruction paradigm relies on the reasoning ability of the LLM to infer high-level semantic information from advertising attributes. The quality of the generated summaries may therefore depend on the capability of the underlying language model. 
In addition, UniSID is currently evaluated primarily in advertising and recommendation scenarios, and its generalization to other SID-based generative modeling tasks remains to be further explored.

\begin{figure*}[h]
\begin{tcolorbox}[
    colback=black!10, 
    colframe=black!70, 
    boxrule=1pt, 
    arc=5mm, 
    boxsep=2mm, 
    left=5mm, 
    right=5mm, 
    top=5mm, 
    bottom=4mm,
    width=\textwidth, 
    nobeforeafter, 
    colupper=black 
]
{\small\sffamily 
\textbf{Instruction:} 

This is an advertisement creative. Please generate a concise summary of the advertisement based on the image and the following information.

Requirements:
1. The summary must include the advertised object (e.g., product or service), the industry, and the first-level category.

2. If key information such as the target audience, product selling points, or promotion strategy can be inferred from the image or text, briefly include them in a one-sentence summary.

3. The output format should be:
   The advertised content is {advertised object}, the industry is {industry}, and the first-level category is {first-level category}. {One-sentence summary (only include core information from the advertisement image and text, describing the main message, selling points, target audience, promotion strategy, etc. Do not describe image details or product specifications. The image should be summarized at a high level, without mentioning model numbers, parameters, or OCR text unless explicitly stated in the provided text.)}
   
4. The summary should be concise and precise, and avoid speculation beyond the advertisement content.

Example:
For an advertisement selling a floral skirt with the title "Lowest price ever — buy now!", the summary can be:
The advertised content is a floral skirt, the industry is general e-commerce, and the first-level category is women's clothing. The advertisement shows a model wearing a floral skirt and attracts customers through a low-price promotion. (The summary is limited to content explicitly shown in the advertisement.)

Below is the information of the advertisement:

} 
\end{tcolorbox}
\caption{Prompt example of the ad attributes summary.}
\label{fig:prompt}
\end{figure*}

\begin{table*}[htbp]
\centering
\caption{More Cases on the Ad-60W dataset. Attributes information extracted from raw advertising data is highlighted in bold, while latent high-level semantics derived via summary-and-reconstruction are marked in red.}
\renewcommand{\arraystretch}{1.3} 

\newcolumntype{L}{>{\raggedright\arraybackslash}m{0.31\textwidth}} 
\scalebox{0.8}{
\begin{tabularx}{\textwidth}{L|L|L}
\toprule
\rowcolor{gray!10} \centering \textbf{Advertisement Attributes} & \centering \textbf{Summary Results} & \centering \textbf{Reconstruction Results} \tabularnewline \midrule

\centering
\vspace{5pt}
\includegraphics[width=0.8\linewidth]{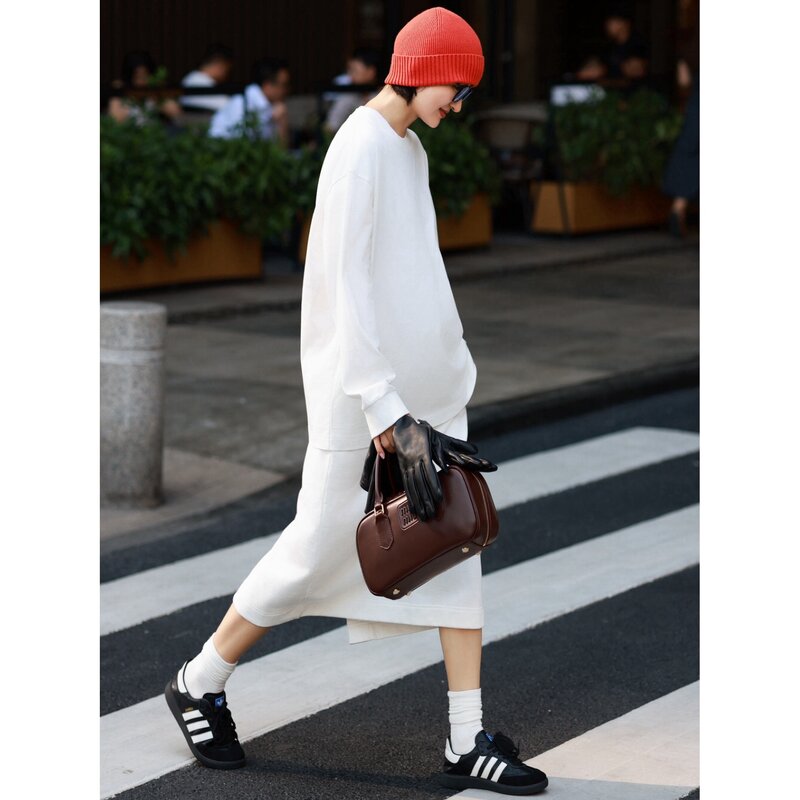} \par
\vspace{5pt}
\begin{flushleft}
\small
{Title:} Versatile Base T-shirt, perfect fit, a practical must-have! \\
{Industry:} General E-commerce \\
{Level 1:} Apparel \& Accessories \\
{Level 2:} Women's Wear \\
{Level 3:} T-shirts
\end{flushleft}
& 
This advertisement sells a versatile base T-shirt. The industry is \textbf{general e-commerce}, and the primary category is \textbf{apparel and accessories}. The ad showcases a street-style look with a model wearing a \textbf{loose white T-shirt paired with fashionable accessories}, highlighting its \textcolor{red}{\textbf{versatile and practical features to attract female consumers pursuing a simple and comfortable style}}.
& 
This advertisement sells a white versatile base T-shirt. The industry is \textbf{general e-commerce}, and the primary category is \textbf{apparel and accessories}. The ad showcases a casual look with a {\textbf{model wearing a white T-shirt paired with black shoes}}, highlighting its \textcolor{red}{\textbf{comfortable and versatile features for female consumers pursuing a simple and fashionable style}}. \tabularnewline \midrule

\centering
\vspace{5pt}
\includegraphics[width=0.85\linewidth]{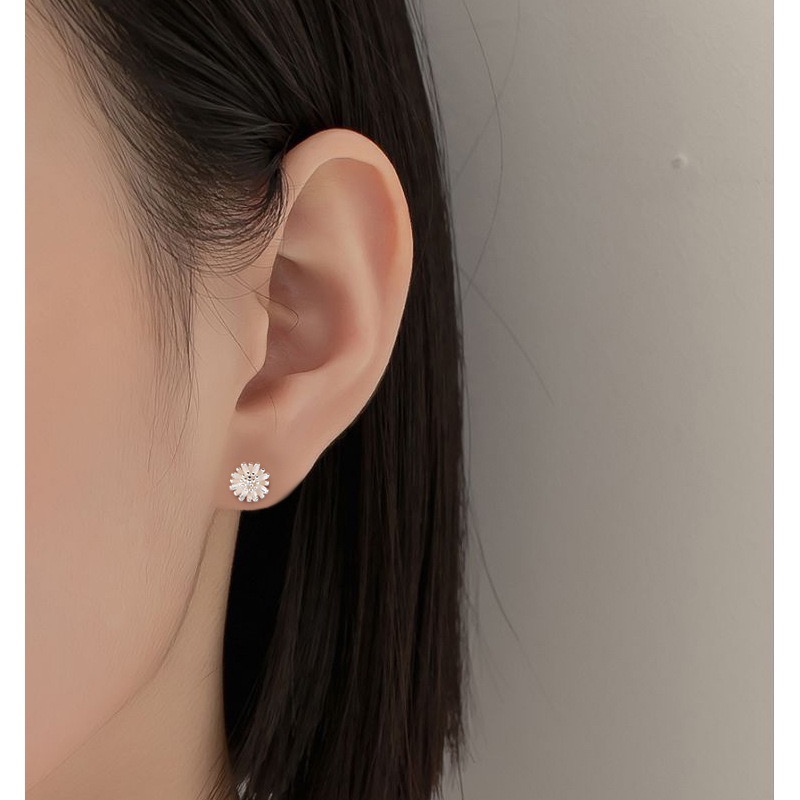} \par 
\vspace{5pt}
\begin{flushleft}
\small
{Title:} Niche Flower Stud Earrings, become an exquisite goddess in summer! S999 Sterling Silver minimalist flower studs, perfect for summer wear. \\
{Industry:} General E-commerce \\
{Level 1:} Jewelry, Jade,  Watches \\
{Level 2:} Silver Jewelry \\
{Level 3:} Ear Jewelry
\end{flushleft}
& 
This advertisement sells S999 sterling silver minimalist flower stud earrings. The industry is \textbf{general e-commerce}, and the primary category is \textbf{jewelry, jade, and watches}. The ad showcases the \textbf{wearing effect of the earrings}, highlighting their \textcolor{red}{\textbf{niche, exquisite, and summer-appropriate}} characteristics to attract \textcolor{red}{\textbf{female consumers pursuing a simple and elegant style}}.
& 
This advertisement sells silver flower stud earrings. The industry is \textbf{general e-commerce}, and the primary category is \textbf{jewelry, jade, and watches}. The ad showcases \textbf{exquisite silver flower stud earrings}, highlighting their \textcolor{red}{\textbf{elegant and fashionable wearing effect}} and emphasizing their \textcolor{red} {\textbf{suitability for daily wear}} to attract \textcolor{red}{\textbf{female consumers pursuing a simple and generous style}}. \tabularnewline \bottomrule

\end{tabularx}}
\label{table:cases}
\end{table*}

\end{document}